\documentclass[12pt,eqsecnum,preprint]{aastex}
\begin{document}
\title{Fast Wave Polarization, Charge Horizons and the Time Evolution of Force-Free Magnetospheres}
\author{Brian Punsly}
\affil{4014 Emerald Street No.116, Torrance CA, USA 90503}
\email{brian.m.punsly@boeing.com or brian.punsly@gte.net}

\begin{abstract} Numerical simulations of force-free, degenerate (ffde)
  pulsar and black hole magnetospheres are often based on  1-D characteristics. In particular,
   the plasma wave polarizations that can be propagated along the 1-D characteristics
   determine the time evolution of the entire system. There are two sets of characteristics,
   corresponding to the fast and Alfven modes. The fast wave is generally considered to be
   a transverse light wave, however recently it has been claimed that light-like fast waves
   can transport a longitudinal electric polarization, $E_{\parallel}$, at the speed of light. The implication is quite profound if true,
   namely that the wrong information has been propagated along the fast characteristics in
   all previous simulations of force-free magnetospheres. It is shown in this Letter that
   the light-like fast waves must be transverse and previous simulations are valid. This
   result is demonstrated by means of a fundamental physical principle (associated with
   the fact that particles cannot flow faster than the speed of light), there exists
   a charge horizon in ffde magnetospheres. It is shown that the Alfven critical surfaces
   in a ffde magnetosphere are both charge and particle horizons, i.e. one way  membranes that
   do not permit traversal by charges nor particles anti-directed to the bulk flow.
   Since the propagation of a discontinuous change in $E_{\parallel}$ requires a physical
   surface charge on the wave-face, it is also a one-way membrane for longitudinally
   polarized waves. Besides justifying previous ffde simulations this result also invalidates
previous claims that fast waves can radiate $E_{\parallel}$ from
the event horizon of a black hole.
\end{abstract}
\section{Introduction} The zero mass, cold limit of MHD (magnetohydrodynamics) known as ffde
is useful for studying tenuous plasmas in strong magnetic fields that might occur in pulsars,
black holes or gamma ray bursts \citep{bla02}. Recently, great progress has been made in
pulsar ffde simulations, \citet{spi03}, and in ffde simulations of black hole magnetospheres,
 \citet{kom01,kom04,uzd04}. To varying degrees each of these simulations require knowledge
 of the 1-D characteristics of the ffde system in order to time evolve the magnetosphere.
 Specifically, the polarization properties of the waves determine the changes in the fields
 that can be propagated at the appropriate speed along a particular characteristic direction.
 The 1-D characteristics represent local plane wave (no transverse variation of the field
 values along the wave face) solutions that can be pieced together on a mesh to produced
 higher dimensional wave structures. Thus, it is crucial that the 1-D characteristics and
 the corresponding local plane waves are well determined before utilizing the method of
 characteristics to time evolve a magnetosphere. There are two plasma modes in the system,
 the fast mode and the Alfven mode. It has been previously shown that a light-like fast wave
 has a transverse electric polarization \citep{bla02,uch97, kom02}. However, recently in
 \citet{lev04}, it was claimed that light-like plane wave solutions can attain a
 longitudinal polarization ($\mathbf{k}\cdot\mathbf{E}\neq 0$, where $\mathbf{k}$ is the
 propagation vector) in a rotating magnetosphere, a case previously considered in
 \citet{kom02}. The polarization information of \citet{kom02} was used to piece together
 2-D simulations of black hole magnetospheres by means of a Riemann solver
  \citep{kom01,kom04}. If the the claim of \citet{lev04} is correct then the simulations
  of \citet{kom01,kom04} are based on the wrong information being propagated along the
  characteristics. This Letter tries to resolve this controversy by studying the global
  physical properties of a ffde magnetosphere as opposed to debating over mathematical errors.
  In particular, we find that the Alfven critical surface is a  one-way membrane for the flow
  of charge, a charge horizon. A similar result was previously shown in \citet{kom04}. It is demonstrated that the
  propagation of a discontinuity in $E_{\parallel}$ requires a physical surface charge
  on the wave-face by Gauss' Law. Hence discontinuities
  in $E_{\parallel}$ (step waves) cannot traverse the charge horizon, anti-directed to the
  flow. It also precludes the possibility of information on $E_{\parallel}$ propagating away from the event
horizon. Finally, the association of $E_{\parallel}$ with the
Alfven wave yields a nice physical interpretation of the Alfven
critical surface as the charge horizon.
  \par The Letter is structured as follows. First of all, the
  maximum velocity of a particle is determined in ffde in section
  2. With this concept, it is straightforward to define the charge
  horizon in section 3, with an example in a black hole
  magnetosphere presented in section 4. In section 5, it is shown mathematically that
  any propagating discontinuity in $E_{\parallel}$ must travel at the
  Alfven wave speed, thus it cannot travel isotropically at the speed of light. Physically, by
  Gauss' law the wavefront is a propagating surface charge, hence it cannot cross the charge horizon
  anti-directed to the flow.
\section{Force-Free Magnetospheric Dynamics}
 The force-free conditions are given by the following relationships, written covariantly in
 terms of the Maxwell field strength tensor, $F^{\mu\nu}$, and four-current
density, $J^{\mu}$, as well as in component form,
\begin{eqnarray}
&& F^{\mu\nu}J_{\nu}=0\;,\quad\rho_{e}\mathbf{E} +\frac{\mathbf{J}\times \mathbf{B}}{c} = 0\;,
\end{eqnarray}
with the implied constraints,
\begin{eqnarray}
&& F^{\mu\nu}F_{\mu\nu}>0\:,\quad \quad \ast F^{\mu\nu}F_{\mu\nu}=0\:.
\end{eqnarray}
 Since the field is degenerate and
magnetic by (2.2), there exists a time-like frame at each point of
space-time in which $\mathbf{E}$ vanishes \citep{kom01}. This is
known as a proper frame (not necessarily a coordinate frame).
Denote the magnetic field in the proper frame as $\mathbf{b}$.
\par It is insightful to decompose a cold MHD plasma into two fluids associated with the two species
of charge and take the zero mass density limit to attain ffde. The
4-current density is expressed in terms of the 4-velocities of the
two fluids, $J^{\mu}=-en_{+}u^{\mu}_{+}+en_{-}u^{\mu}_{-}$. The
momentum equations of the two fluids are:
\begin{eqnarray}
&& n_{\pm}m_{\pm}
        \frac{d}{d t}\mathbf{u}_{\pm} =
        \mp\frac{en_{\pm}}{c}\left(\mathbf{E} +
        \frac{\mathbf{u}_{\pm}\times \mathbf{B}}{c}\right)\;.
\end{eqnarray}
Adding the two momentum equations and taking the limit of
$nm\rightarrow 0$, yields the force-free condition (2.1). Notice
that in the same limit, (2.3)  implies that all particles flow
parallel to $\mathbf{b}$, otherwise the resulting $\mathbf{E}$ in
the frame of the particles would drive large cross-field
(nonforce-free) currents.
\par The maximum velocity that a particle can attain in the force-free limit is the speed of
light along $\mathbf{b}$. Transforming this velocity from the proper frame to a general
orthonormal frame by a local Lorentz boost $\mathbf{E}\times\mathbf{B}/B^{2}$ yields the
maximum three velocity of a force-free particle in a local frame \citet{lig75,bla02},
\begin{eqnarray}
&&\mathbf{V}_{max}=c\frac{\mathbf{E}\times\mathbf{B}+\left(B^{2}-E^{2}\right)^{1/2}\mathbf{B}}{B^{2}}\;.
\end{eqnarray}
\section{The Charge Horizon}It is useful to define the proper frame known as the corotating
frame of a ffde magnetosphere. For such a frame to be meaningful,
the poloidal magnetic flux should be slowly changing (so as to
make the azimuthal electric field, $E^{\phi}$, negligibly small by
Faraday's Law) as in the flat spacetime, rotating, stationary
magnetosphere of \citet{lev04}. The corotating frame is attained
by a local azimuthal boost, $v^{\phi}_{F}$,  from an any
orthonormal frame at fixed poloidal coordinate in an axisymmetric
magnetosphere, $ v^{\phi}_{F}=E^{P}/B^{P}$, where the superscript
P represents the poloidal component. If there is a non-vanishing
$E^{\phi}$ then a poloidal boost is required in addition to an
azimuthal boost, which greatly complicates the algebra. At a light
cylinder,  $E^{P}/B^{P}=\pm 1$ by definition. The light cylinder
is also an Alfven critical surface in ffde which is a one-way
membrane for Alfven waves, no Alfven waves can cross the surface
anti-directed to the bulk flow \citep{bla02}. Using (2.4), the
poloidal component of the maximum velocity of a force-free
particle is
\begin{eqnarray} && V^{P}_{max}= \pm\frac{-E^{P}B^{\phi}+\sqrt{B^{2}-E^{2}}B^{P}}{B^{2}}\;,
\end{eqnarray}
where the plus (minus) sign determines the direction of the bulk
flow (i.e., it is positive in a pulsar wind and negative near the
event horizon of a black hole). Outside of the light cylinder,
$\mid E^{P}/B^{P} \mid > 1$ and inside the light cylinder $\mid
E^{P}/B^{P} \mid < 1$. Consequently by (3.1), in the rotating
magnetosphere of \citet{lev04} or in a pulsar, $V^{P}_{max}=0$ at
the light cylinder,  $V^{P}_{max}>0$ inside the light cylinder and
$V^{P}_{max}< 0$ outside of the light cylinder. Thus, no charges
or particles can cross the Alfven critical surface (light
cylinder) anti-directed to the flow - the Alfven surface is a
charge and particle horizon in an axisymmetric ffde magnetosphere.
By (2.4), at the light cylinder, the anti-directed particles
spiral azimuthally at the speed of light with no poloidal
velocity.
\section {Black Hole Charge Horizons} As expected by the equivalence
principle, it is straightforward to show that the concept of a
charge horizon can be extended to curved space-time by means of an
explicit example. We consider the case of the magnetosphere of a
rotating black hole described by the Kerr metric. There are two
light cylinders and hence two Alfven critical surfaces in a black
hole magnetosphere, one is the standard outer light cylinder of
pulsar physics that was discussed in section 3 and the other is an
inner light cylinder that is associated with the dragging of
inertial frames in the ergosphere \citep{bla02}. Consider the
initial state of the simulation of \citet{kom01}. In ingoing
Kerr-Schild (K-S, hereafter) coordinates, the initial state has
only one component of $F_{\mu\nu}$ that represents a purely radial
magnetic field. From (2.3), as $nm\rightarrow 0$, one has
$F_{\mu\nu}u^{\nu}_{\pm}=0$. Evaluating this in K-S coordinates,
tells us that particles moving along $\mathbf{b}$ satisfy the
condition $d\tilde{\phi}=0$. A constant K-S azimuthal coordinate,
$d\tilde{\phi}=0$ transforms to the constraint $d\phi=-(a/\Delta)
dr$ in Boyer-Lindquist (B-L, hereafter) coordinates, where
$\Delta\equiv r^{2}-2Mr+a^{2}$. Note that this implies a strong
toroidal magnetic field near the event horizon in B-L coordinates
\citep{pun04}. The $d\tilde{\phi}=0$ condition defines the
particle trajectories that are restricted to the field lines, a
requirement of ffde that was derived in section 2. As discussed in
\citet{pun04}, the inner light cylinder is at the outer boundary
of the ergosphere, the stationary limit in the initial state (the
initial state has zero field line angular velocity as viewed from
asymptotic infinity, $\Omega_{_{F}}=0$). In order to evaluate
(2.4) and (3.1), we need to introduce an orthonormal frame so that
the concept of three-velocity is well-posed. The most famous such
frame in the Kerr space-time is the ZAMO frame (see \citet{lig75},
\citet{pun01} and references therein for a review of ZAMOs and the
stationary limit surface). Using the ZAMO field values in the
initial state, found in \citet{pun04}, evaluated at the stationary
limit (inner light cylinder) shows that $V^{P}_{max}=0$ at the
light cylinder, $V^{P}_{max}<0$ inside the light cylinder and
$V^{P}_{max}> 0$ outside of the light cylinder as expected from
section 3.
\par One can get more insight into the Alfven critical condition by
looking at the azimuthal velocity of a particle as measured in the
ZAMO frames, $ c\beta^{\phi}$, that flows outward (as viewed in
the global B-L coordinates) inside of the inner Alfven surface.
First transform the four velocity from B-L coordinates to the ZAMO
frames, then apply this result to trajectories restricted to
$\mathbf{b}$, $d\tilde{\phi}=0$,
\begin{mathletters}
\begin{eqnarray}
&& u^{\phi}=\sqrt{g_{\phi\phi}}\frac{d\phi}{d\tau}-\Omega\sqrt{g_{\phi\phi}}\frac{dt}{d\tau}\;,\\
&& u^{0}=\alpha \frac{dt}{d\tau}\;,\\
&& \beta^{\phi}= \frac{u^{\phi}}{u^{0}}=-\frac{\frac{a}{\Delta}\frac{dr}{dt}-\Omega}{c\alpha}\sqrt{g_{\phi\phi}}\;,\quad \frac{dr}{dt}>0\;,
\end{eqnarray}
\end{mathletters}
where $\tau$ is the proper time of the particle, the angular
velocity of the ZAMOs as viewed from asymptotic infinity,
$\Omega\equiv -g_{\phi t}/g_{\phi\phi}$, is defined in terms of
the metric in B-L coordinates and
$\alpha=\sqrt{\Delta\sin^{2}{\theta}/g_{\phi\phi}}$ is the lapse
function (that describes the redshift of the ZAMO frames as viewed
from asymptotic infinity) that vanishes at the event horizon. By
(4.1), for outgoing trajectories, the maximum permissible
azimuthal \textbf{three-velocity} (corresponding to $dr/dt=0$) is
equal to -c at the stationary limit and decreases without bound as
one approaches the event horizon. As in flat spacetime, an
outgoing particle can spiral endlessly at the Alfven critical
surface, but never flow through it.
\par Consider the ramifications of (4.1c) to a hypothetical charge
that appears to flow outward as seen globally ($dr/dt>0$) that is
inside
  of the inner Alfven critical surface. Since there is a strong
  toroidal magnetic field extant near the horizon in the initial state,
  the physical charges flow on spiral trajectories. Inside of the Alfen critical surface,
  the resulting spiral flow velocity of a globally outgoing force-free charge is
  necessarily much larger than the speed of light in order that the poloidal projection
  of the flow velocity be sufficient to escape black hole gravity. Thus any out-flowing
  charge in this region is acausal.
\section{Longitudinally Polarized Fast Wave} According to
eqns. (8) and (11) of \citet{lev04}, a fast wave can propagate at
the speed of light with $E_{\parallel}$. The causality question is
whether there is a light-like wave that can propagate changes in
$E_{\parallel}$. It is useful to look at abrupt discontinuities in
order to resolve this question for two reasons. First of all, the
simulations of \citet{kom01,kom04} evolve by means of a Riemann
solver that propagates discontinuities that make changes in the
field variables. Secondly, the step wave analysis of MHD waves is
a method that was specifically developed to provide clarity to the
causal evolution of waves (since they are launched by a well
defined piston) in the classic work of \citet{kap66}. In the WKB
notation of \citet{lev04}, the oscillatory nature wave functions
is given by
\begin{equation}
e^{i(\mathbf{\nabla}\psi(\mathbf{x})\cdot\mathbf{x}-\omega
t)}=e^{-ik(\mathbf{x})(v_{_{F}}t-X_{n})}\;,
\end{equation}where $v_{_{F}}$ is the fast wave phase speed and
$X_{n}$ is a local coordinate along $\mathbf{k}$. For
non-dispersive light-like waves, the step function essentially
contains information on all of the oscillatory modes since it is
given by the Fourier composition,
\begin{equation}
\Theta(ct-X_{n})=-\frac{1}{2\pi
i}\int_{-\infty}^{+\infty}\frac{e^{-ik(ct-X_{n})}dk}{k+i\epsilon}\;,
\end{equation} where $\epsilon$ is an
arbitrarily small positive number in the usual sense.
\par If the light-like $E_{\parallel}$ conjecture is physical then
it must result from an analysis of abrupt discontinuities in cold
MHD in the limit of zero plasma mass density. We will evaluate the
discontinuities in the rest frame of the wavefront and take the
limit of zero mass. Technically there is no light-like frame of
reference, nevertheless our results will be well-defined in the
limiting process for which time-like wavefronts exist. MHD
discontinuities are solved for by considering the continuity of
the stress-energy tensor, $T^{\mu\nu}$, across the wavefront. Let
x be the local normal coordinate to the wavefront and the upstream
magnetic field, $\mathbf{B}$ is in the x-y plane and z lies in the
wavefront surface. All we need consider to get the desired results
is mass conservation, the frozen-in condition, which are
\begin{eqnarray}
&& nu^{x}u^{0}=\mathrm{constant}\;,\;
\mathbf{E}+\frac{1}{c}\mathbf{v}\times\mathbf{B}=0\;,
\end{eqnarray}
the antisymmetric components of Maxwell's equations
\begin{eqnarray}
&&
F_{\alpha\beta_{;}\gamma}+F_{\gamma\alpha_{;}\beta}+F_{\beta\gamma_{;}\alpha}=0\;,
\end{eqnarray}
 and the continuity of one component of the stress-energy tensor,
$T^{xz}$,
\begin{eqnarray}
&& n\mu u^{x}u^{z}-\frac{1}{4\pi}(E^{x}E^{z} +
B^{x}_{u}B^{z})=0\;,
\end{eqnarray} in which all of the quantities are evaluated
downstream unless there is a subscript "u" and $\mu$ is the
specific enthalpy (note that by (5.3), $E^{x}_{u}=0$). The
continuity of $B_{x}$ used in (5.5) follows from the
$\nabla\cdot\mathbf{B}=0$ condition in (5.4). A tremendous
simplification occurs in (5.4) at the step wavefront, since to
lowest order, all the singular terms must cancel (surface terms
like delta functions), thus the normal covariant derivative does
not depend on the connection coefficients and one has continuity
of $E^{y}$ and $E^{z}$ at the wavefront, or
\begin{eqnarray}
&&\frac{\partial E^{y}}{\partial x}= \frac{\partial
E^{z}}{\partial x}=0\;,
\end{eqnarray}
across the wavefront. Inserting (5.3) and (5.6) into (5.4), one
gets
\begin{eqnarray}
&&
\left((u^{x})_{u}^{2}\left[\frac{n_{u}}{n}+\frac{(b^{y}_{u})^{2}}{4\pi
n_{u}\mu c^{2}}\right]- \frac{(b^{x}_{u})^{2}}{4\pi
n_{u}\mu}\right)E^{x}=0\;,
\end{eqnarray}
written in terms of the field in the plasma rest frame,
$\mathbf{b}$. For $n_{u}=n$, $u^{x}_{u}$ is the Alfven or
intermediate wave speed in MHD \citep{pun01}. Taking the limit of
zero mass density, (5.7) has two solutions,
\begin{eqnarray}v\equiv\frac{u^{x}_{u}}{u^{0}_{u}}= \mathrm{c}\,
\cos{\theta}\;,\mathrm{or}\quad E^{x}=0\;,
\end{eqnarray}
where $\theta$ is the angle between $\mathbf{b}$ and the wave
normal in the plasma rest frame and $v$ is the force-free value of
the Alfven speed in a proper frame \citep{pun03}. Thus by (5.8), a
force-free discontinuity travels at the Alfven speed (which is not
the speed of light in general), or it carries no $E_{\parallel}$.
\section {Discussion} It was shown in the last section that
force-free fast waves cannot propagate changes in $E_{\parallel}$,
in agreement with \citet{kom02}. A discontinuity in
$E_{\parallel}$ requires a surface charge density, $\sigma=
E_{\parallel}/4\pi$, by Gauss' law. There is no precursor to
$\sigma$ and it is the source of the wave. In fact, it can be
shown that the motion of $\sigma$ along the magnetic field line in
the proper frame (the surface current) and Maxwell's equations
entirely determine the fields in the downstream state
\citep{pun03}. The Alfven surface in ffde is a one-way membrane
for the physical charges that comprise $\sigma$ and therefore for
waves carrying $E_{\parallel}$. Furthermore, the existence of the
charge horizon and the inherent charge of Alfven discontinuities
provides a fundamental physical description of the Alfven critical
surface in ffde.
\par How do we explain the results of \citet{lev04}? That
calculation was as attempt to collect higher order terms in a WKB
approximation. The danger in doing this is that one must be
rigorous in making sure that all terms of higher order are
retained in all equations from the beginning. The inconsistent
equation is the normal component of Ampere's law, $F^{x\mu}_{\quad
;\mu} =4\pi J^{x}/c$. It was not justified to ignore the
transverse derivative, $\mathbf{\nabla}_{tr}$ of the tangential
magnetic field, $B_{tan}$. From eqns. (8) and (11) of
\citet{lev04}, $\partial E^{x}/\partial t \sim (w/c)(\delta
E_{2}/kR)=\delta E_{2}/R$ in the notation of \citet{lev04} in
which R is a cylindrical radius. Therefore $\partial
E^{x}/\partial t\sim \mathbf{\nabla}_{tr}\times \delta B_{tan}$
(the change in the poloidal electric field in a force-free wind
also results in a change in the toroidal magnetic field, in fact
the two field components are approximately equal in the asymptotic
field zone beyond the Alfven critical surface \citep{ogu03}). In
particular the normal component of Ampere's Law changes from
$\partial E^{x}/\partial t = 4\pi J^{x}$ in \cite{lev04} to
${\nabla}_{tr}\times \delta B_{tan}=4\pi J^{x}/c$ in the full 2-D
discontinuity calculation. The failure of \citet{lev04} is that as
the wavelength increases, plane waves are more and more inaccurate
representations of the plasma waves in a relativistic
magnetosphere. One must find more exact 2-D solutions, with
variations transverse to the propagation vector to learn more than
the standard WKB approximation. The 2-D solutions should reveal
any dispersive effects of the fast waves that result from boundary
conditions as is the case in a vacuum waveguide or a plasma-filled
waveguide \citep{pun01}. The axisymmetric 2-D fast wave
discontinuities in an ffde black hole magnetospheres were
determined in \citet{pun04}. The oscillatory 2-D fast waves in a
black hole magnetosphere are linear combinations of spin-weighted
spheroidal harmonics convolved with a radial function that is a
solution of an extremely complicated differential equation
\citep{teu73,pun01}.

\end{document}